# Competition between orbital effects, Pauli limiting, and Fulde-Ferrell-Larkin-Ovchinnikov states in 2D transition metal dichalcogenide superconductors

Chang-woo Cho[1], Cheuk Yin Ng[1], Mahmoud Abdel-Hafiez[2], Alexander N. Vasiliev[3,4,5], Dmitriy A. Chareev[4,5,6], A. G. Lebed[7,8] and Rolf Lortz[1,*]

[1]Department of Physics, The Hong Kong University of Science and Technology, Clear Water Bay, Kowloon, Hong Kong

[2]Department of Physics and Astronomy, Uppsala University, Uppsala SE-75120, Sweden

[3]Department of Low Temperature Physics and Superconductivity, Lomonosov Moscow State University, Moscow 119991, Russia

[4]Quantum Functional Materials Laboratory, National University of Science and Technology "MISiS" Moscow 119049, Russia

[5]Ural Federal University, Ekaterinburg 620002, Russia

[6]Institute of Experimental Mineralogy, RAS, Chernogolovka, Moscow Region 142432. Russia

[7]Department of Physics, University of Arizona, 1118 E. 4-th Street, Tucson, AZ 85721, USA

[8]Landau Institute for Theoretical Physics, RAS, 2 Kosygina Street, Moscow 117334, Russia

[*]Correspondence to Rolf Lortz (lortz@ust.hk).

We compare the upper critical field of bulk single-crystalline samples of the two intrinsic transition metal dichalcogenide (TMD) superconductors, 2H-NbSe$_2$ and 2H-NbS$_2$, in high magnetic fields where their layer structure is aligned strictly parallel and perpendicular to the field, using magnetic torque experiments and a high-precision piezo-rotary positioner. While both superconductors show that orbital effects still have a significant impact when the layer structure is aligned parallel to the field, the upper critical field of NbS$_2$ rises above the Pauli limiting field and forms a Fulde-Ferrell-Larkin-Ovchinnikov (FFLO) state, while orbital effects suppress superconductivity in NbSe$_2$ just below the Pauli limit. From the out-of-plane anisotropies, the coherence length perpendicular to the layers of 31 Å in NbSe$_2$ is much larger than the interlayer distance, leading to a significant orbital effect suppressing superconductivity before the Pauli limit is reached, in contrast to the more 2D NbS$_2$.

Type-II spin-singlet superconductors have upper critical fields $B_{c2}$, usually dominated by the orbital limit for superconductivity [1], when the superconducting screening currents reach a value at which the Cooper pairs break apart. Strongly anisotropic layered superconductors can be an exception if the magnetic field is applied strictly parallel to the layer structure [2-17]. If the coupling between the layers is weak, the orbital effect is suppressed and the orbital limit can exceed the Pauli limit for superconductivity [18,19]. At the Pauli limiting field, the energy of Zeeman splitting between the two electrons forming the Cooper pair reaches a value at which pair formation is abruptly suppressed and the normal state is restored via a discontinuous first-order phase transition. In contrast, the orbital limit of type-II superconductor induces a continuous second-order transition to the normal state.

Transition metal dichalcogenide (TMD) superconductors such as 2H-NbSe$_2$, 2H-NbS$_2$ and 2H-TaS$_2$ are strongly anisotropic layered superconductors in which two-dimensional planes are weakly coupled by van der Waals forces. TMD materials can be exfoliated down to monolayer thickness and have been the focus of recent research due to their wide range of unique electronic properties with high potential for technological applications [20-27]. In their 2D form, they have aroused great interest due to the discovery of Ising superconductivity, which allows them to exceed the Pauli limit of superconductivity [23,24]. However, there is also another mechanism that allows superconductors to maintain their superconducting state above the Pauli limit, which is the formation of a Fulde-Ferrell-Larkin-Ovchinnikov (FFLO) state [28,29]. It arises when Cooper pairs acquire finite centre-of-mass momentum, leading to a spatially modulated order parameter that stabilizes the superconducting condensate. Superconductivity can then exist well beyond the theoretical Pauli limit in the form of a pair density wave state [30]. FFLO states have been reported in layered organic superconductors [5-15], and in the form of the Q-phase in the layered heavy-fermion superconductor CeCoIn$_5$ [2-4]. Other examples include the iron-based superconductors KFe$_2$As$_2$ [16], and FeSe [31]. The formation of the FFLO state requires that the superconductor is in the clean limit and the mean-free path $\ell$ of the electrons exceeds the coherence length $\xi$ [32]. This is the case in TMD superconductors, where, for example, $\ell$ = 320 Å in 2H-NbSe$_2$ exceeding the very short $\xi$ by a factor of 10 [33].

We recently reported the observation of an FFLO state in 2H-NbS$_2$ bulk samples when we applied the magnetic field exactly parallel to its layer structure with an accuracy of one millidegree using a piezo rotary stage in combination with magnetic torque, specific heat and thermal expansion experiments [17]. In this article, we report magnetic torque measurements for 2H-NbSe$_2$ for precisely parallel alignment of the magnetic field with respect to the layer structure and compare the magnetic field phase diagram with the one of 2H-NbS$_2$.

## Results

In layered superconductors with strong 2D character, it is essential to achieve a strictly parallel field orientation to minimize the orbital effect and find the maximum critical field $B_{c2\parallel}$. We achieved this by first minimizing $\tau$ at a fixed temperature and field (0.3 K, 4T) by rotating the sample stepwise through the parallel orientation [7,16,17]. In this way, we could approximately estimate the parallel alignment. We then repeated the field scans at tiny angular variations of 0.1° or less until we found torque data with minimum amplitude, minimal hysteresis loop opening and maximum $B_{c2}$. This is consistent with parallel alignment. Fig. 1a shows data for NbS$_2$ where the hysteresis loop almost disappears when a maximum $B_{c2}$ indicates that the parallel orientation has been reached. NbSe$_2$, for which data near parallel orientation are shown in Fig. 1b at two very small angles of $\theta$ = 0.1° and 0°, does not show this dramatic change in torque behaviour near parallel alignment. Even when $B_{c2}$ is maximized at 11.2 T, the hysteresis loop remains wide open. Note that, like bulk magnetization, the torque is composed of a reversible thermodynamic contribution and an irreversible contribution resulting from current loops induced by flux pinning. Macroscopic screening currents can flow in the out-of-plane direction only if there is a certain coupling among the planes. Here, hysteresis indicates sufficient coupling between the planes for such screening currents to flow and contribute to torque, whereas in NbS$_2$ they almost vanish for strictly parallel fields. In the weak coupling BCS limit the Pauli limit for superconductivity can be estimated as $B_P = 1.84\, T_c$. Note that the $B_{c2\parallel}$ value of NbSe$_2$ is slightly smaller than the theoretical Pauli limit at 13 T. In contrast, as previously reported [17], $B_{c2\parallel}$ of NbSe$_2$ clearly exceeds the Pauli limit at ~10 T.

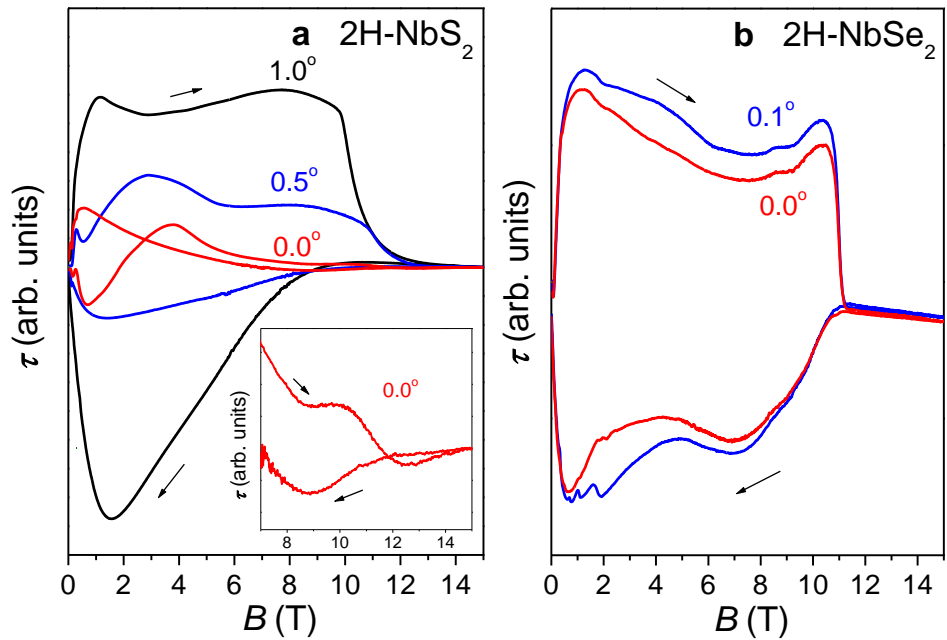

**Fig. 1 a** Magnetic torque measured at $T = 0.3$ K at $\theta = 1.0°$, $0.5°$ and $0.0°$ (field parallel to basal plane) for 2H-NbS$_2$. The exactly parallel alignment can be seen from the closure of the hysteresis loop and the maximum $B_{c2}$. The inset shows a magnified view of the high-field region for the $\theta = 0.0°$ data. Note that $B_{c2\parallel}$ cannot be reached in the field range up to 15 T for this angle. **b** Similar data for 2H-NbSe$_2$ for $\theta = 0.1°$ and $0.0°$, where the hysteresis loop remains wide open even at such small angles, unlike NbS$_2$.

Fig. 2a shows the magnetic torque for the exactly parallel field orientation at different fixed temperatures. The torque signal increases rapidly at small fields, reaches a maximum, and then initially decreases. At 5 K the torque continuously disappears at $B_{c2}$ near 4 T, except for a tiny anomaly indicating enhanced flux pinning known as the peak effect [35]. At lower temperatures, the torque begins to rise again at higher fields, forming a pronounced peak before abruptly dropping to a small background value in the normal state at $B_{c2}$. Such a sharp decay of the screening currents is usually observed in Pauli-limited superconductors and associated with a first-order nature of the $B_{c2\parallel}$ transition triggered by the Pauli paramagnetic effect [7,16,17]. However, NbSe$_2$ does not reach the Pauli limit, even if it comes close to it. Therefore, it remains unclear whether this strong decay is already triggered by the influence of the Pauli paramagnetic effect, which certainly plays a role near the theoretical Pauli limit. A complete hysteresis loop recorded at 0.3 K is shown in Fig. 1b. It shows how the shielding currents drop abruptly as $B_{c2}$ is approached, but build up much more continuously as the field is reduced. Fig. 2b shows the magnetic torque in various fixed fields measured under field-cooled conditions during temperature sweeps across the superconducting transition.

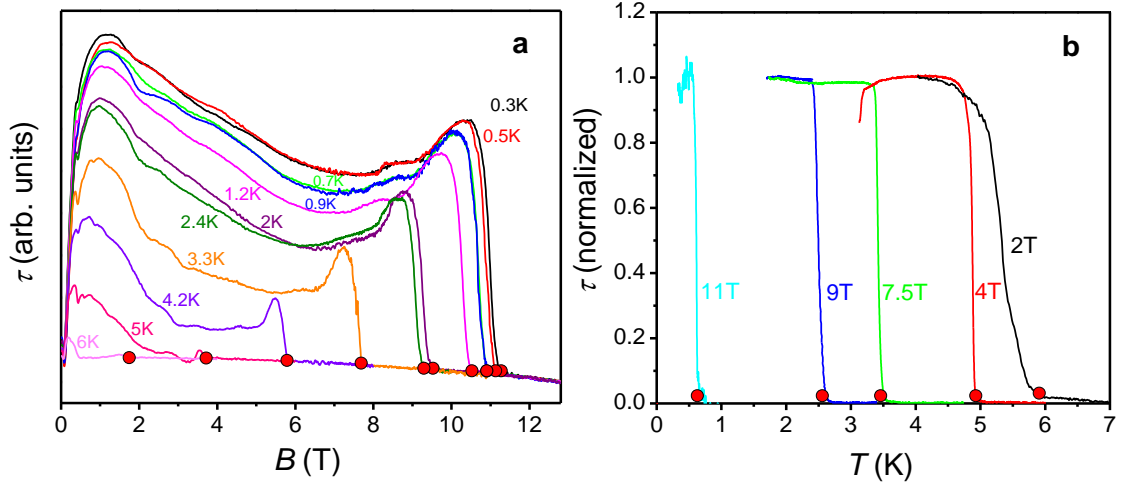

**Fig. 2** Magnetic torque data in parallel fields of 2H-NbSe$_2$. **a** Magnetic torque magnitude $\tau(B)$ measured at fixed temperatures as a function of increasing magnetic field applied strictly parallel to the basal plane ($\theta = 0°$). **b** Magnetic torque $\tau(T)$ measured in fixed parallel fields as a function of temperature. The data have been normalized for clarity. The additional circles mark the critical fields $B_{c2}(T)$ (a) or critical temperatures $T_c(B)$ (b) to be added in a magnetic field vs. temperature phase diagram (Fig. 3).

|  | $T_c$ | $B_P$ | $B_{c2\parallel}$ exp. | $B_{c2\perp}$ exp. | $B_{c2\parallel}$ orb. WHH/GL | $B_{c2\perp}$ orb. WHH/GL | $\xi_{0\parallel}$ | $\xi_{0\perp}$ | $\xi_{0\parallel}/\xi_{0\perp}$ | $l(H)/d$ |
|---|---|---|---|---|---|---|---|---|---|---|
| NbSe$_2$ | 7.1 K | 13 T | 11.2 T | 5 T | 11.2 T \ 15 T | 5 T / 6.2 T | 71 Å | 31 Å | 2.29 | 7.5 |
| NbS$_2$ | 5.5 K | 10 T | >>15 T | 1.8 T | 24 T \ 33 T | 1.8 T / 2.2 T | 120 Å | 8 Å | 15 | 0.8 |

**Table 1.** Relevant superconducting parameters for 2H-NbSe$_2$ and 2H-NbS$_2$: critical temperature $T_c$, theoretical Pauli limit $B_P = 1.84 T_c$ in the weak coupling limit, experimentally determined upper critical field in parallel ($B_{c2\parallel}$) and perpendicular fields ($B_{c2\perp}$), extrapolated upper critical fields according to the WHH and GL models in parallel ($B_{c2\parallel}$) and perpendicular fields ($B_{c2\perp}$), in-plane ($\xi_{0\parallel}$) and out-of-plane coherence length ($\xi_{0\perp}$) at $T=0$, ratio between in-plane and out-of-plane coherence length ($\xi_{0\parallel}/\xi_{0\perp}$), ratio between electron quasi-classical motion length and layer distance ($l(H)/d$). For details see text.

## Discussion

In Fig. 3, we plot the extracted $B_{c2}$ values of NbSe$_2$ for both the parallel ($B_{c2\parallel}$) and perpendicular ($B_{c2\perp}$) field directions with respect to the layer structure in a magnetic phase diagram, compared to the one of NbS$_2$ we reported in Ref. [17] (see Table 1 for an overview of the relevant superconducting parameters). $B_{c2\parallel}$ was taken as the upper limit at which the torque deviates from the small normal state background. Additional DC magnetization data were used to determine $B_{c2\perp}$, for which accurate alignment is much less critical and which agree perfectly with Ref. [36]. The theoretical Pauli limits of 10 T (NbS$_2$) and 13 T (NbSe$_2$) for both compounds are indicated by the dotted horizontal lines. The dashed lines represent fits with the standard Werthamer–Helfand–Hohenberg (WHH) model [37] that illustrate the expected temperature dependence of the orbital limit for superconductivity if there were no Pauli paramagnetic effect. The WHH model describes well the temperature dependence of $B_{c2}$ for NbSe$_2$ in both field directions. $B_{c2\parallel}$ comes close to the Pauli limit but does not reach it. For a layered superconductor, the anisotropy is comparatively weak, with $B_{c2\parallel}$ =11.2T being just over twice the value of $B_{c2\perp}$ =5T. Note that our $B_{c2}$ values at low temperatures differ slightly from those found in the literature [33], which is due to the fact that these previously reported values are based on extrapolations of critical field values or anisotropies at higher temperatures, which may be affected by the positive curvature of the $B_{c2}$ line near the zero field $T_c$.

In contrast, $B_{c2\parallel}$ for NbS$_2$ clearly exceeds the Pauli limit of 10 T. Above the Pauli limit, the $B_{c2\parallel}$ curve begins to deviate from the WHH model, indicating the sudden effect of the Pauli paramagnetic effect, before rising again. This upturn, together with the thermodynamic signature of an additional phase transition occurring near 10 T in the specific heat and magnetostriction, has been interpreted as consequence of the formation of the FFLO state [17], which allows the superconductor to maintain its state above the Pauli limit. The anisotropic coherence lengths can be deducted from the linear Ginzburg-Landau (GL) fits (dotted lines), which provide an estimation of the GL orbital limit [38] for superconductivity in parallel fields at $B_{c2\parallel}$ = 33 T, as opposed to a low $B_{c2\perp}$ = 2.2 T.

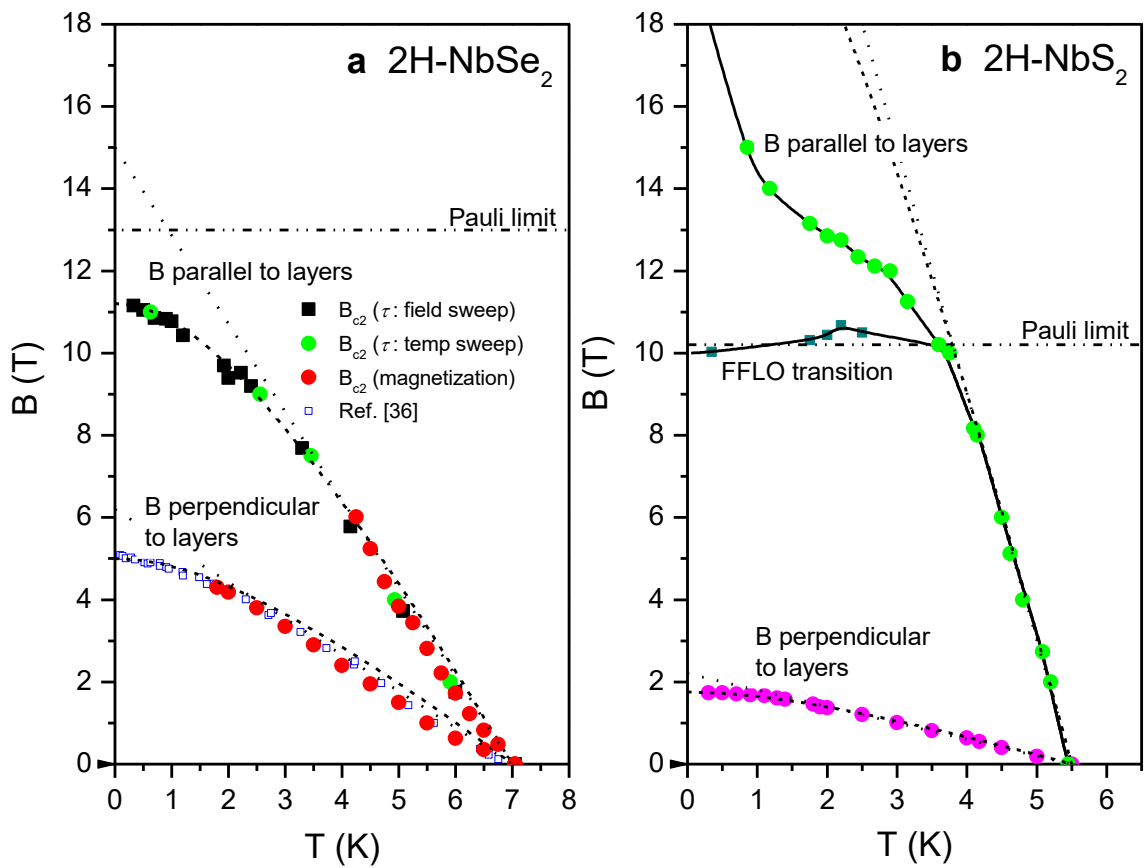

**Fig. 3** Magnetic phase diagram of 2H-NbSe$_2$ (a) and 2H-NbS$_2$ (b) [17] in fields applied strictly parallel and perpendicular to the layer structure. The horizontal lines mark the Pauli limits at 13 and 10 T, respectively. The data in (a) were obtained from magnetic torque measurements supplemented below 7 T by DC magnetization measurements. Data from Ref. [36] were added for the perpendicular direction. It shows the upper critical field lines for both field orientations. The data in (b) were compiled from magnetic torque, specific heat and thermal expansion data (see Ref. 17 for details). The olive squares in (a) mark a phase transition boundary within the superconducting state of NbS$_2$ attributed to the transition to an FFLO state in the high field region [17]. The dashed and dotted lines are fits with the standard Werthamer–Helfand–Hohenberg (WHH) model and the Ginzburg-Landau (GL) model, respectively, providing estimates of the orbital limits for superconductivity.

Note that the NbSe$_2$ $B_{c2}$ data deviate somewhat from the GL and WHH models at high temperatures, as the data exhibit some concave temperature dependence. This has been observed previously [33,36] and is likely the effect of fluctuations near $T_c$ that are enhanced by the low-dimensionality and the relatively high $T_c$ value of NbSe$_2$.

From the anisotropy of the $B_{c2}$ values at low temperatures, we can derive the perpendicular and parallel coherence lengths $\xi_{0\perp}$ = 31 Å and $\xi_{0\parallel}$ = 71 Å for NbSe$_2$ from Eq. 1 and 2, respectively. If we use the GL linear extrapolated orbital limit in parallel fields (Fig. 3a) as input for the orbital upper critical field in NbS$_2$, we obtain $\xi_{0\perp}$ = 8 Å and $\xi_{0\parallel}$ = 120 Å.

$$B_{c2\parallel}(0K) = \Phi_0/(2\pi\xi_{0\perp}\xi_{0\parallel}) \quad (1)$$

$$B_{c2\perp}(0K) = \Phi_0/(2\pi\xi_{0\parallel}^2) \quad (2)$$

Our comparison of the magnetic phase diagrams NbSe$_2$ and NbS$_2$ demonstrates that NbSe$_2$ is much less 2D than NbS$_2$ with a much weaker anisotropy $\xi_{0\parallel}/\xi_{0\perp}$ = 2.29 in contrast to the large value of NbS$_2$ $\xi_{0\parallel}/\xi_{0\perp}$ = 15. This explains why the 3D orbital effects in NbSe$_2$ are very strong, as shown by the rapid suppression of the critical temperature in parallel magnetic fields driven by the orbital effect. The difference in the phase diagram for high magnetic fields is certainly related to the ratio of $\xi_{0\perp}$ to the interlayer spacing, $d$ [39,40]. For both compounds, the $c$-axis lattice parameter in the out-of-plane direction is about $d$ = 18 Å [41], and it is evident that for NbSe$_2$ $\xi_{0\perp}$ = 31 Å causes strong coupling of the superconducting layers with significant Meissner currents in the out-of-plane direction, suppressing superconductivity by the orbital effect before the Pauli limit is reached at 13 T. For NbS$_2$, the much shorter $\xi_{0\perp}$ = 8 Å brings the superconductor to the borderline of the 2D limit where the layers begin to decouple, and the orbital effect is significantly weakened, allowing the Pauli limit to be reached and the FFLO state to be formed.

A more rigorous theory of the competition between the orbital effects and the FFLO phase has been recently suggested in Ref. [42], where the quantum effects of electron motion in a parallel magnetic field in layered superconductors are considered. It was shown that the destructive 3D orbital effects against superconductivity destroy the FFLO phase if $l(H)/d \gg 1$, and that the FFLO phase occurs if $l(H)/d < 1$, where $l(H)$ is the magnitude of the electron quasi-classical motion in a direction perpendicular to the conducting layers. By applying of the results of Ref. [42] to the experimental data obtained in this paper, we find that, indeed, in NbSe$_2$ $l(H)/d$ = 7.5, whereas in NbS$_2$ $l(H)/d$ = 0.8, which explains the destruction of the FFLO phase by the 3D Meissner currents in NbSe$_2$ and its stability in NbS$_2$.

In conclusion, the two TMD superconductors 2H-NbSe$_2$ and 2H-NbS$_2$ are sister compounds with very similar superconducting characteristics. However, the larger Se ions in 2H-NbSe$_2$ significantly reduce the out-of-plane anisotropy, so they exhibit different magnetic field vs. temperature phase diagrams in strictly parallel fields. In 2H-NbSe$_2$, the weaker anisotropy enhances the coupling between the planes, allowing sufficiently strong Meissner shielding currents to flow between the layers so that the orbital effect becomes the leading mechanism for suppressing superconductivity near, but still below, the Pauli limit for superconductivity. In contrast, the smaller S ions in 2H-NbS$_2$ suppress orbital effects sufficiently for the upper critical field $B_{c2}$ to reach the Pauli limit at ~4 K. At lower temperatures, the formation of an unusual FFLO state leads to a strong enhancement of $B_{c2}$.

Most previous work on FFLO states has been performed on layered organic superconductors, which require very high magnetic fields on the order of 20 T to study, which are only available in large scale high magnetic field facilities. 2H-NbS$_2$ allows the study of the FFLO state in a field and temperature range accessible in a standard cryogenic laboratory. The availability of two very similar TMD superconducting compounds provides a unique opportunity to determine the parameters required for an FFLO state and encourages experiments in which the interlayer coupling is varied by either ion substitution or high pressure.

## Methods

### Sample Preparation

High quality single crystals of transition metal dichalcogenides were prepared by an evaporation method in a fused silica ampoule. To achieve supersaturation, the ampoule was subjected to a controlled temperature gradient of 100 – 70 °C, with the hot side maintained at 850 – 400 °C. Crystallization was initiated over a period of ~3 weeks by slowly cooling the multicomponent flux to reduce the solubility of the components. The chalcogenide feed was introduced into the hot part of the reaction vessel, where it dissolved in a growth medium in the form of an eutectic CsCl-KCl-NaCl salt melt mixture, from where it migrated to the cold side to crystallize. A detailed characterization of the process can be found in Ref. 34. The resulting high-quality single-crystalline platelets with optically flat surfaces on both sides were cut into a square shape of ~1 mm side length.

### Magnetic Torque Measurements

The magnetic torque $\boldsymbol{\tau}$ is a vector quantity directly related to the anisotropic DC magnetization: $\boldsymbol{\tau} = \mathbf{M} \times \mathbf{B}$, where $\mathbf{B}$ is the applied magnetic field. It was measured using a capacitive cantilever technique [7,16,17]. The sample was attached to a circular plate with a diameter of 5 mm at the end of the cantilever leg, forming one of the plates of a parallel plate capacitor. The plates are electrically insulated by a sapphire disk to allow reversible measurements as a function of field and temperature. The sensor was mounted on a piezo rotary stage in a $^3$He probe of a 15 T magnet cryostat, so that the layered structure of the sample could be aligned with millidegree accuracy with respect to the field direction. Field or temperature sweeps were performed at a rate of 0.1 - 0.5 T/min and 0.04 K/min, respectively. To achieve perfectly parallel field alignment, it is essential that the crystals are mounted absolutely flat on the cantilever. Since our sample was only about 100 microns thin, this was ensured by using the adhesive force of highly diluted GE 7031 varnish to firmly attach the sample to the flat, polished cantilever plate. Capacitance was measured using a General Radio 1615-A capacitance bridge and a SR830 digital lock-in amplifier. Note that even with parallel alignment, $\boldsymbol{\tau}$ usually does not completely vanish because of quadrupole moments [7,16,17]. All data present the magnitude $\tau$ of the torque.

## Data availability

The data that support the findings of this study are available from the corresponding author upon reasonable request.

## Acknowledgments
We thank U. Lampe for technical assistance. This work was supported by grants from the Research Grants Council of the Hong Kong Special Administrative Region, China (GRF-16302018, GRF-16303820, C6025-19G-A, SBI17SC14). M. A. H. acknowledge the financial support from the Swedish Research Council (VR) under project No. 2018-05393. Support by the P220 program of Government of Russia through the project 075-15-2021-604 is acknowledged.

## Author contributions
This work was initiated by R.L.; C.w.C. and C.Y.N. carried out the magnetic torque experiments; the single crystal sample was provided by A.N.V, D.A.C and M.A.H.; A.G.L. provided the theoretical support. The manuscript was prepared by R.L. and all authors were involved in discussions and contributed to the manuscript.

## Competing financial interests
The authors declare no competing financial interests.